\documentclass[12pt]{article}
\usepackage{psfig,epsfig}
\begin{document}
\def\ee{e^+e^-}
\def\ttbar{t\overline t}
\def\eett{$\ee \rightarrow \ttbar$~}
\def\stsp{$\sin\theta\,\sin\phi$~}
\def\stcp{$\sin\theta\,\cos\phi$~}
\def\lq{leptoquark~}
\def\lqs{leptoquarks~}
\def\beq{\begin{equation}}
\def\eeq{\end{equation}}
\def\bea{\begin{eqnarray}}
\def\eea{\end{eqnarray}}
\def\gve{g_V^e}
\def\gae{g_A^e}
\def\gvt{g_V^t}
\def\gat{g_A^t}
\begin{flushright}
hep-ph/0408083
\end{flushright}
\begin{center}
{\Large \bf \boldmath 
Transverse beam polarization and limits on leptoquark couplings in
$e^+e^- \to t \overline t$}
\vskip .5cm
{ Saurabh D. Rindani}
\vskip .2cm
{\it Theory Group,
Physical Research Laboratory}\\
{\it Navrangpura, Ahmedabad 380 009, India}

\vskip 1cm
{\bf Abstract}
\end{center}
\begin{quote}
It is shown that if electron and positron beams at a linear collider are
transversely polarized, azimuthal asymmetries of the final-state top quark in \eett can be
used to probe a combination of couplings of left and right chiralities in a
scalar leptoquark model. 
The CP-conserving azimuthal asymmetry
would be a sensitive test of the chirality violating couplings. 
A linear 
collider operating at $\sqrt{s}=500$ GeV and having transverse polarizations of
80\% and 60\% respectively for the $e^-$ and $e^+$ beams, can put a limit of
the order of $0.025$ 
on the product of the left and right chirality \lq
couplings 
(in units of the electromagnetic coupling constant),
with a \lq mass of 1 TeV and 
for an integrated luminosity of 500 fb$^{-1}$.
The CP-violating azimuthal asymmetry, which would
provide a direct test of CP-violating phases in leptoquark couplings, 
can be constrained to the same level of accuracy.
 However, this limit is uninteresting in view of the much better
indirect limit from the 
electric dipole moment of the electron. 
\end{quote}
\vskip 1cm
\section{Introduction}

An $\ee$ linear collider operating at a centre-of-mass (cm) energy of several
hundred GeV will offer an opportunity to make precision measurement of the
properties of the electroweak gauge bosons, top quarks, Higgs bosons, and also
to constrain new physics. 
Linear colliders are expected to  have the option of
longitudinally polarized beams, which can help to improve the sensitivity of
these measurements and reduce background in the search for new physics. It has
been realized that spin rotators can be used to convert the longitudinal beam
polarization to transverse polarization. This has inspired studies which
investigate the role of transverse polarization in constraining new
physics \cite{tp, ba}, though these studies are yet far from being exhaustive.

It was pointed out recently \cite{ba} (see also \cite{hikasa} for an earlier
discussion) that transverse polarization 
can play a unique role in isolating chirality violating couplings, to which
processes with longitudinally polarized beams are not sensitive. 
The interference of new chirality-violating contributions with the
chirality-conserving standard model (SM) couplings give rise to terms in the
angular distribution proportional to \stcp and 
\stsp, where $\theta$ and $\phi$ are the
polar and azimuthal angles of a final-state particle. Chirality conserving new
couplings, on the other hand, produce interference contributions 
proportional to 
$\sin^2\theta\,\cos2\phi$ and $\sin^2\theta\,\sin2\phi$. Chirality-violating
contributions do not interfere with the chirality-conserving SM 
contribution with
unpolarized or longitudinally polarized beams when the electron mass is
neglected. Hence transverse polarization would enable measurement of
chirality-violating couplings through the azimuthal distributions.

A general discussion of azimuthal distributions and asymmetries arising with 
transverse
beam polarization in the context
of CP violation was presented in \cite{ba}, and illustrated by means of the
process \eett in the presence of general contact interactions. It is difficult to come by
models where chirality violating couplings which produce the specific azimuthal distributions 
mentioned above are present at low orders of perturbation.  In this paper,
we examine a specific model where chirality-violating couplings occur at tree
level, viz., a scalar leptoquark model. In this model, there is an $SU(2)_L$
doublet of scalar leptoquarks, which couples only to first-generation leptons 
and
third-generation quarks. Since couplings of \lqs to the third generation quarks
are relatively weakly constrained, their effect in \eett is expected to be
non-negligible. This model has been chosen mainly for purposes of illustration 
of the ideas in \cite{ba}, and we find that transverse polarization can
indeed be used to put direct constraints on such a model. It would be 
interesting to look for the azimuthal asymmetries described here if a
future linear collider can be equipped with transversely polarized
beams.

We allow \lq couplings of both left and right chiralities, and also allow them
to be complex. Thus, the possibility of CP violation is kept open. We then show
how azimuthal asymmetries of the top quarks in the process \eett with
transversely polarized beams can be used to measure the phases of the couplings.

\section{The model} 

We now go to the details of the model. We assume that SM with its gauge group
$SU(2)_L\times U(1)\times SU(3)_C$ is extended by a  multiplet $\phi$ of scalar
leptoquarks transforming according to the representation $(\underline{2},
-\frac{7}{6},\underline{3}^*)$ of the gauge group. Assuming $\phi$ to couple
only to first generation leptons and third generation quarks, its couplings to
the fermions can be written as 
\beq\label{lag}
{\cal L}_{\phi} = h_{2L} \overline{l}_L u_R \phi + h_{2R}^*\overline{q}_L i
\tau_2 e_R \phi^* + {\rm H.c.},
\eeq 
where
\beq
l_L \equiv \left( \begin{array}{c} \nu_e \\ e \end{array} \right)_L,\;\;
q_L \equiv \left( \begin{array}{c} t \\b \end{array} \right)_L,
\eeq
are left-handed doublets. The representation for $\phi$ has been chosen so that
it can contribute to the process
$$ e^-_Le^+_L \to t_R\overline{t}_R $$
and 
$$ e^-_Re^+_R \to t_L\overline{t}_L $$
by a $t$-channel exchange. In SM, on the other hand, $s$-channel exchange of
$\gamma$ and $Z$ 
contributes only to 
$$ e^-_Le^+_R \to t\overline{t} $$
and 
$$ e^-_Re^+_L \to t\overline{t}. $$
In the above, the  subscripts $L$, $R$ denote chiralities. Since we will neglect
the electron mass, these will be identical to helicities so far as the $e^+$ and
$e^-$ are concerned.

It is possible to choose a scalar \lq multiplet transforming as $(\underline{1},
{1 \over 3}, \underline{3}^*)$, which satisfies the conditions stated above. The
corresponding couplings would be fermion-number violating. We refer the reader
to \cite{buch} for a general discussion of \lq models, and to 
\cite{tana} for a brief review of quantum numbers. 
However, the results
so far as azimuthal distributions are concerned would be analogous to the case
we treat here.

\section{The process \eett}

The amplitude for the process 
\beq\label{process}
e^-(p_1,s_1) + e^+ (p_2,s_2) \to t(k_1) + \overline{t}(k_2),
\eeq
with the couplings of ${\cal L}_{\phi}$, in addition to SM couplings, is 
\beq
M = M_1 + M_2,
\eeq
where the SM contribution is
\bea
M_1& =& e \left[ \frac{2}{3}\overline{u}(k_1)\gamma^{\mu} v(k_2) 
\frac{1}{q^2}\overline{v}(p_2,s_2)
\gamma_{\mu} u(p_1,s_1)
+
\overline{u}(k_1)\gamma^{\mu}(g_V^t-g_A^t \gamma_5)v(k_2)
\right. \nonumber\\
&&
\left. \times
 \left(-g_{\mu\nu} + 
{q_{\mu}q_\nu
\over m_Z^2} \right)\frac{1}{q^2-m_Z^2}
\overline{v}(p_2,s_2)\gamma^\nu(g_V^e - g_A^e \gamma_5)u(p_1,s_1)\right],
\eea
where $q=p_1+p_2=k_1+k_2$. The contribution $M_2$ coming from $t$-channel \lq
exchange is 
\beq
M_2 = e \!\left[ \overline{u}(k_1)(g_LP_L + g_RP_R) u(p_1,s_1) \frac{1}{t-M^2} 
\overline{v}(p_2,s_2) (g_L^*P_R + g_R^*P_L) v(k_2) \right],
\eeq
where we have used the notation $h_{2L,R} = e g_{L,R}$, $P_{L,R} = {1\over 2}
(1\mp \gamma_5)$, and $M$ is the mass of the leptoquark exchanged.  
The couplings to $Z$ of the fermions are given by
\beq
\begin{array}{ll}
g_V^e = -\frac{1}{4} + \sin^2\theta_W, & g_A^e = -\frac{1}{4},
\end{array}
\eeq
\beq
\begin{array}{ll}
g_V^t = \frac{1}{4} - \frac{2}{3}\sin^2\theta_W, & g_A^t = \frac{1}{4}.
\end{array}
\eeq

We assume transverse polarizations $P_1$ and $P_2$ of the $e^-$ and
$e^+$ beams, respectively, which are assumed to be parallel to each other, apart
from a possible sign. 
We can then use simple Dirac algebra to
obtain the cross section as the sum of the SM contribution $\sigma_{\rm SM}$,
the pure \lq contribution $\sigma_{\rm LQ}$, and the contribution
$\sigma_{\rm int}$ from the interference of the \lq contribution with the SM
contribution. We can write the differential cross section as
\beq
\frac{d\sigma}{d\Omega}
= \frac{d\sigma_{\rm SM}}{d\Omega} + 
\frac{d\sigma_{\rm LQ}}{d\Omega} + \frac{d\sigma_{\rm
int}}{d\Omega}.
\eeq
Here the SM differential cross section is itself the sum of the $\gamma$
contribution, the $Z$ contribution, and the $\gamma Z$ interference
contribution:
\beq\label{diffSM}
\frac{d\sigma_{\rm SM}}{d\Omega} = \frac{d\sigma_{\rm SM}^{\gamma}}{d\Omega}+
\frac{d\sigma_{\rm SM}^Z}{d\Omega} + \frac{d\sigma_{\rm SM}^{\gamma
Z}}{d\Omega},
\eeq
where 
\beq
\frac{d\sigma_{\rm SM}^{\gamma}}{d\Omega} = \frac{\alpha^2\beta}{3s} \left[
2- (1+P_1P_2)\beta^2\sin^2\theta + 2 P_1P_2 \beta^2\sin^2\theta\cos^2\phi
\right],
\eeq
\bea
\frac{d\sigma_{\rm SM}^Z}{d\Omega}& =& \frac{3\alpha^2\beta s}{4(s-m_Z^2)^2} 
\left[ (g_V^{t\,2} + g_A^{t\,2})\left\{  (g_V^{e\,2} + g_A^{e\,2}) - 
(g_V^{e\,2} -
g_A^{e\,2}) P_1P_2 \beta^2\right\} \right. \nonumber \\
&&\left. + (g_V^{t\,2} - g_A^{t\,2}) (g_V^{e\,2} + g_A^{e\,2}) (1-\beta^2) + 8 g_V^tg_A^t
g_V^eg_A^e \beta \cos\theta \right. \nonumber \\
&& \left. + (g_V^{t\,2} + g_A^{t\,2})\left\{  (g_V^{e\,2} + g_A^{e\,2}) + (g_V^{e\,2} - 
g_A^{e\,2}) P_1P_2 \right\} \beta^2 \cos^2\theta \right. \nonumber \\
&& \left. +2 (g_V^{t\,2} + g_A^{t\,2}) (g_V^{e\,2} - 
g_A^{e\,2}) P_1P_2 \beta^2 \sin^2\theta \cos^2 \phi \right],
\eea
\bea
\frac{d\sigma_{\rm SM}^{\gamma Z}}{d\Omega}&=&-\frac{\alpha^2\beta}{s-m_Z^2}
\left[g_V^e g_V^t \left\{2 - (1 + P_1P_2)\beta^2\sin^2\theta \right\}
\right. \nonumber \\
&& + \left. 2 g_A^e g_A^t \beta \cos^2\theta + 2 g_V^e g_V^t
P_1P_2\beta^2\sin^2\theta \cos^2\phi \right].
\eea

The pure \lq contribution is given by
\beq
\frac{d\sigma_{\rm LQ}}{d\Omega}=\frac{3\alpha^2\beta s}{64
(t-M^2)^2} \left( \vert g_L \vert^2 + \vert g_R \vert^2 \right)^2 (1-\beta
\cos\theta)^2.
\eeq

The contribution from the interference between the \lq and the SM $\gamma$ and
$Z$ diagrams is, respectively, 
\beq
\frac{d\sigma_{\rm int}^{\gamma}}{d\Omega}=\frac{\alpha^2\beta}{8(t-M^2)} 
\left( \vert g_L \vert^2 + \vert g_R \vert^2 \right)\!\!\left\{\!(P_1P_2  
  \cos 2 \phi  - 1 )\beta^2 \sin^2 \theta +2- 2\beta \cos \theta)\!\right\},
\eeq
and
\bea
\frac{d\sigma_{\rm int}^{Z}}{d\Omega}&\!\!\!\!=\!\!\!\!&
\!\!- \frac{3\alpha^2\beta s}{16 (s-m_Z^2)
(t-M^2)}\!\! \left[ \vert g_R \vert^2 \!\left\{ 
\!P_1P_2 \beta^2 (\gve-\gae) (\gvt-\gat) \sin^2 \theta \cos 2 \phi 
\right. \right. \nonumber \\ 
&&\!\!\!\!\!\!\!+\!\! \left.\left.
(\gve + \gae ) ( \gvt + \gat )
(1-\beta^2) 
+ (\gve+\gae) (\gvt-\gat ) (1 - \beta\cos\theta)^2 
\right\} \right. \nonumber \\
&&\!\!\!\!\!\!\! + \left. \vert g_L \vert^2 \left\{
P_1P_2 \beta^2 (\gve+\gae) (\gvt+\gat) \sin^2 \theta \cos 2 \phi
\right. \right. \nonumber \\
&&\!\!\!\!\!\!\!+\!\! \left.\left.
(\gve - \gae ) ( \gvt - \gat )
(1-\beta^2)
+ (\gve-\gae) (\gvt+\gat ) (1 - \beta\cos\theta)^2
\right\} \right. \nonumber \\
&& \!\!\!\!\!\!\! + \left. 4m_t \sqrt{s} \beta (\gve \gat + \gae \gvt)
\sin\theta \left\{
(P_1+P_2){\rm Re}\left(g_Rg_L^*\right)\cos \phi \right.\right. \nonumber\\
&&\!\!\!\!\!\!\! +\left. \left.
 (P_1-P_2){\rm Im}\left(g_Rg_L^*\right)\sin \phi \right\}
\right] .
\eea
In the above, $t=(p_1-k_1)^2 = m_t^2 - \frac{s}{2}(1-\beta \cos\theta)$.

It can be seen from these equations that the interference term between the SM
$Z$ contribution and the \lq contribution contains terms proportional to \stcp
and \stsp, which are linear in $P_1$ and $P_2$, and are proportional
respectively to the real and imaginary parts of $g_Rg_L^*$. Both these require
the simultaneous presence of couplings of both chiralities. The term containing
\stsp is a measure of CP violation, and is nonzero only if $g_L$ and $g_R$ are
relatively complex. These terms do not need both $e^-$ and $e^+$ beams to be
polarized. There are also terms in the differential cross section proportional
to $\sin^2\theta\sin 2\phi$ and $\sin^2\theta\cos 2\phi$, which are proportional
to $P_1P_2$, and to $\vert g_L\vert^2$ or $\vert g_R\vert^2$. They are present
even if leptoquark coupling of only one chirality is present, but require both
$e^-$ and $e^+$ beams to be polarized. These absolute values of chiral couplings 
are also the ones that can be studied using longitudinal polarization, because in 
that case the interference between different chirality contributions vanish in the 
limit of vanishing electron mass.

The chirality violating terms can be isolated by studying the following
azimuthal asymmetries, where we assume $\theta$ to be integrated over with a
cut-off $\theta_0$ in the forward and backward directions:
\beq
A_1(\theta_0) = { 1\over\sigma(\theta_0)} \int_{-\cos\theta_0}^{\cos\theta_0}
	d\cos\theta\left[ \int_0^\pi d\phi - \int_\pi^{2\pi} d\phi \right]
\frac{d\sigma}{d\Omega},
\eeq
\beq
A_2(\theta_0) = { 1\over\sigma(\theta_0)} \int_{-\cos\theta_0}^{\cos\theta_0}
	d\cos\theta\left[ \int_{-\pi /2}^{\pi /2} d\phi - \int_{\pi /2}^{3\pi /2}
d\phi \right]
\frac{d\sigma}{d\Omega},
\eeq
where 
\beq
\sigma(\theta_0) = \int_{-\cos\theta_0}^{\cos\theta_0}
d\cos\theta
\int_0^{2\pi} d\phi
\;\frac{d\sigma}{d\Omega}.
\eeq

The expressions for $A_1$ and $A_2$ in the leading order, where terms 
quartic in the \lq couplings are neglected, are as follows:
\bea\label{a1}
A_1(\theta_0)& =& {1\over\sigma_{\rm SM}(\theta_0)}\,(P_1-P_2)\, \left(\gve\gat +
\gae\gvt \right){\rm Im}(g_Rg_L^*)\frac{6\alpha^2m_t}{s^{3/2}(s-m_Z^2)}
\nonumber \\
&\times &\left[C(\pi - 2 \theta_0) - 2\sqrt{C^2-s^2\beta^2}\tan^{-1} \left(
\frac{\sqrt{C^2-s^2\beta^2}}{C}\cot\theta_0 \right) \right],
\eea
\bea\label{a2}
A_2(\theta_0)& =& {1\over\sigma_{\rm SM}(\theta_0)}\,(P_1+P_2)\, \left(\gve\gat +
\gae\gvt \right){\rm Re}(g_Rg_L^*)\frac{6\alpha^2m_t}{s^{3/2}(s-m_Z^2)} 
\nonumber \\
&\times &\left[C(\pi - 2 \theta_0) - 2\sqrt{C^2-s^2\beta^2}\tan^{-1} \left(
\frac{\sqrt{C^2-s^2\beta^2}}{C}\cot\theta_0 \right) \right],
\eea
where $C=2M^2 - 2m_t^2 + s$, and $\sigma_{\rm SM}(\theta_0)$ is the SM cross
section with the cut-off $\theta_0$, which may be easily evaluated by an
appropriate integration of $d\sigma_{\rm SM}/d\Omega$ in eq. (\ref{diffSM}).
It can be seen from eqs. (\ref{a1}) and (\ref{a2}) that $A_1(\theta_0)$ and
$A_2(\theta_0)$ differ only in the factors 
$(P_1-P_2)\,{\rm Im}(g_Rg_L^*)$ and $(P_1+P_2)\,{\rm Re}(g_Rg_L^*)$.
The two asymmetries, therefore, will have identical dependence on the SM
parameters and kinematic variables.

\begin{figure}[htb]
\centering
\hskip -2cm
\psfig{file=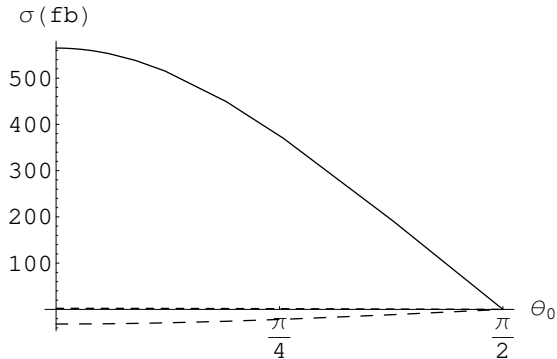,height=6cm}
\caption{The various contributions to the total cross section with a
cut-off $\theta_0$ for values $g_L=g_R=1/\sqrt{2}$, and leptoquark mass $M=1000$
GeV. The solid curve is the SM contribution, the long-dashed curve the 
interference between the SM and \lq contributions, and the 
short-dashed curve is the pure \lq contribution.} 
\end{figure}

\section{Numerical Results}
We now come to the numerical results. For our calculations we use
$\alpha=1/128$, $m_Z=91.1876$ GeV, $m_t =  174$ GeV, and $\sin^2\theta_W =
0.233$. We assume a cm energy of $\sqrt{s} = 500$ GeV,  a linear collider
with $e^-$ polarization 80\%, $e^+$ polarization 60\%, and an integrated
luminosity of 500 fb$^{-1}$.

\begin{figure}[htb!]
\centering
\hskip -2cm
\epsfig{file=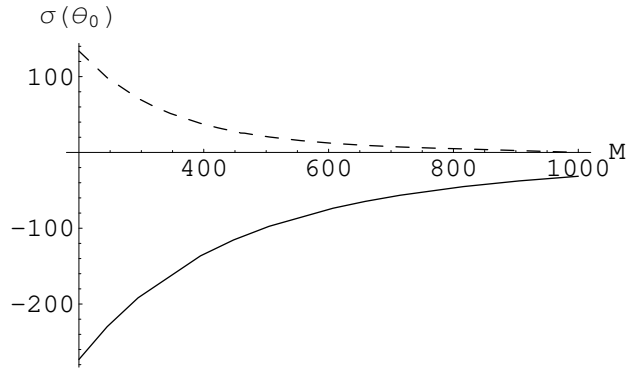,height=6cm}
\caption{The interference (solid curve) and pure \lq contributions (dashed curve)
to the cross
section  as a function of the
leptoquark mass $M$ (in GeV), for a fixed value of cut-off, 
$\theta_0=0.1$ radians. The other parameters are as in Fig.~1} 
\end{figure}

In Fig. 1 we show the different contributions to the total cross section with a
cut-off $\theta_0$ for values $g_L=g_R=1/\sqrt{2}$, and leptoquark mass $M=1000$
GeV. The solid curve is the SM contribution, the long-dashed curve the 
interference between the SM and \lq contributions,
which is approximately the total new-physics contribution, since the pure \lq
contribution (short-dashed curve) is negligible. 
The cross sections in Fig. 1 are
independent of transverse polarization. They show a monotonic decrease with 
$\theta_0$, as expected.

Fig. 2 shows the interference between the SM and \lq amplitudes (solid curve),
 and pure \lq contributions (dashed curve)
 to the cross section  as a function of the
leptoquark mass $M$ (in GeV), for a fixed value of cut-off,
$\theta_0=0.1$ radians. The other parameters are as in Fig. 1. As
expected, the leptoquark contribution decreases with $M$.

\begin{figure}[htb!]
\centering
\epsfig{file=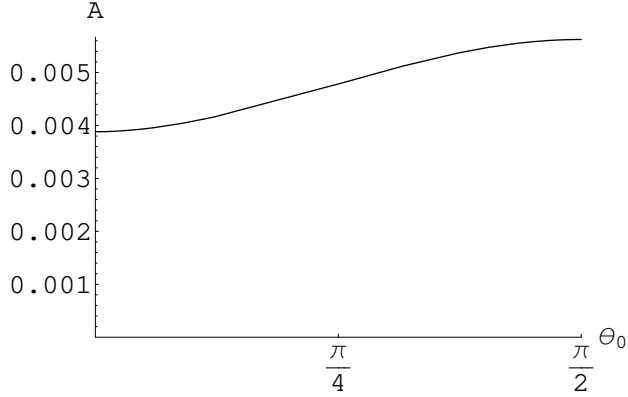,height=6cm}
\caption{
The asymmetry $A_1(\theta_0)$ as a function of $\theta_0$ for
the values $P_1=0.8$, $P_2=-0.6$, $g_L=1/\sqrt{2}$, $g_R=i/\sqrt{2}$ and
$M=1000$ GeV. The same curve also shows $A_2(\theta_0)$ for $g_L=1/\sqrt{2}$,
$g_R=1/\sqrt{2}$, and $P_2=0.6$.}
\end{figure}

Fig. 3 depicts the asymmetry $A_1(\theta_0)$ as a function of $\theta_0$ for
the values $P_1=0.8$, $P_2=-0.6$, $g_L=1/\sqrt{2}$, $g_R=i/\sqrt{2}$ and
$M=1000$ GeV. The values of the couplings correspond to maximal CP violation
in the leptoquark couplings.  The asymmetry is of the order of $4\times
10^{-3}$, and is not very sensitive to the cut-off.

In view of the remark made earlier, Fig. 3 also shows the asymmetry
$A_2(\theta_0)$ for the values $P_1=0.8$, $P_2=0.6$, $g_L=1/\sqrt{2}$,
$g_R=1/\sqrt{2}$ and
$M=1000$ GeV. In this case, there is no CP violation, and the sign of $P_2$ is
chosen positive to maximize the asymmetry. 

\begin{figure}[htb!]
\centering
\epsfig{file=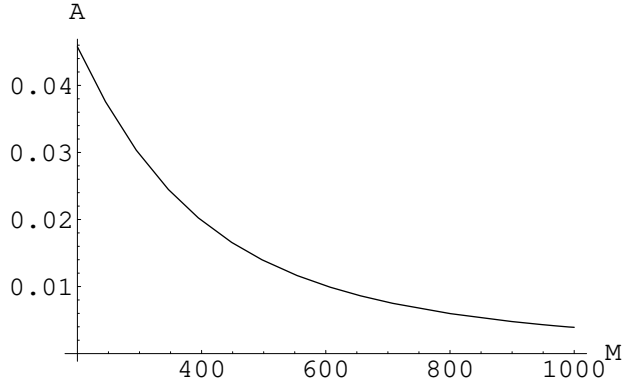,height=6cm}
\caption{
The asymmetry $A_1(\theta_0)$ as a function of 
$M$ (in GeV) for
the values $P_1=0.8$, $P_2=-0.6$, $g_L=1/\sqrt{2}$, $g_R=i/\sqrt{2}$, and
$\theta_0 = 0.1$ radians.
The same curve also shows $A_2(\theta_0)$ for $g_L=1/\sqrt{2}$,
$g_R=1/\sqrt{2}$ and $P_2=0.6$.}
\end{figure}

\begin{figure}[htb]
\centering
\epsfig{file=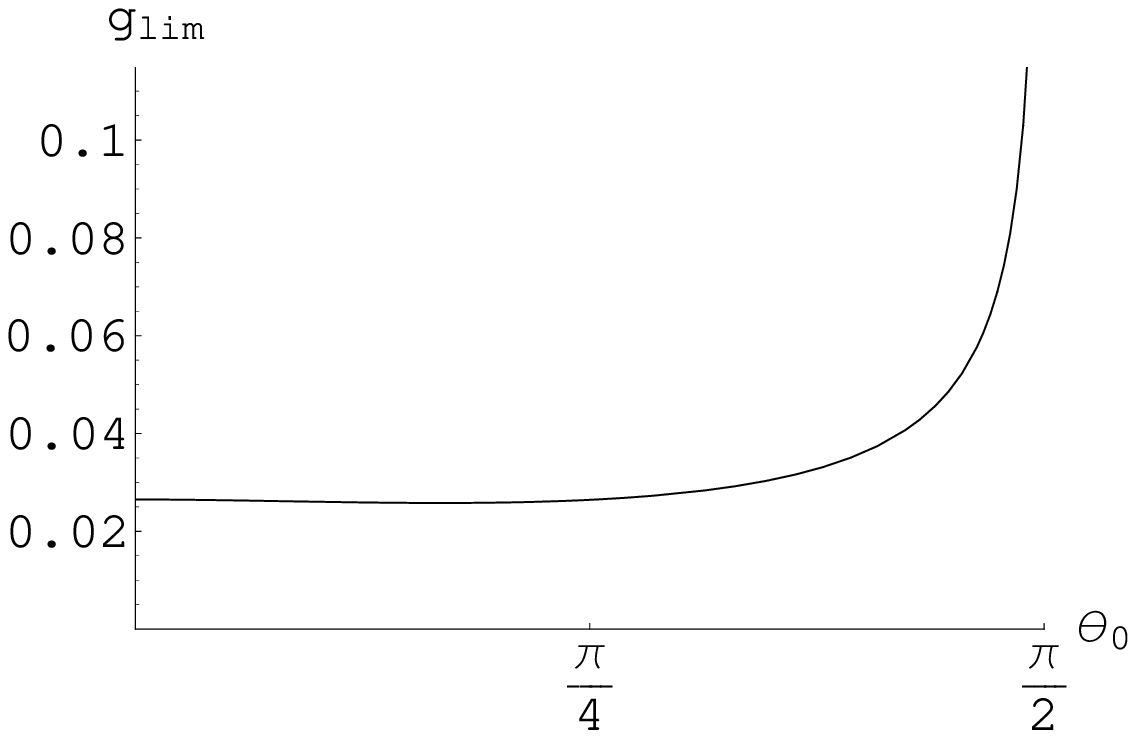,height=6cm}
\caption{
The 90\% CL limit $g_{\rm lim}$ that can be obtained on ${\rm Re} (g_Rg_L^*)$
or ${\rm Im} (g_Rg_L^*)$ respectively from $A_1$ or $A_2$ for an integrated
luminosity of 500 fb$^{-1}$ plotted as a function of $\theta_0$.}
\end{figure}

\begin{figure}[htb!]
\centering
\epsfig{file=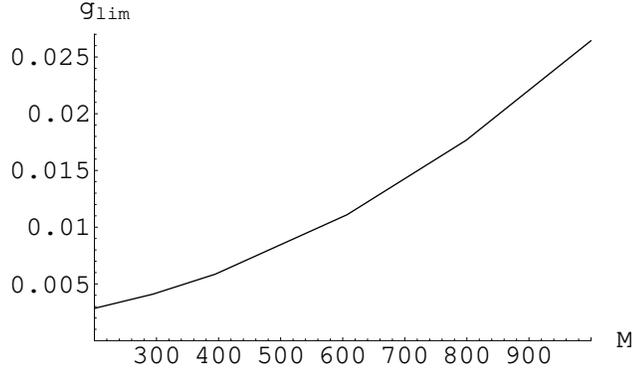,height=6cm}
\caption{
The 90\% CL limit $g_{\rm lim}$ that can be obtained on ${\rm Re} (g_Rg_L^*)$
or ${\rm Im} (g_Rg_L^*)$ respectively from $A_1$ or $A_2$ for an integrated
luminosity of 500 fb$^{-1}$ plotted as a function of $M$ (in GeV).}
\end{figure}

We plot in Fig. 4 the asymmetry $A_1(\theta_0)$ (or $A_2(\theta_0)$  with a
suitable change of parameters) as a function of $M$ for $\theta_0 = 0.1$
radians.

We show in Fig. 5 the 90\% confidence level (CL) limit $g_{\rm lim}$ that can
be put on the combinations ${\rm Im}(g_Rg_L^*)$ (in the maximal CP violation
case) and ${\rm Re}(g_Rg_L^*)$ (in the CP conservation case) for $M=1000$ GeV.
This limit is obtained by equating the asymmetry to 1.64/$\sqrt{N_{\rm SM}}$,
where $N_{\rm SM}$ is the number of SM events, $N_{\rm SM}= \sigma_{\rm SM}
(\theta_0)\,L$, $L$ being the integrated luminosity. It can be seen that the
possible limit $g_{\rm lim}$ on ${\rm Re}(g_Rg_L^*)$ or ${\rm Im}(g_Rg_L^*)$ is
about 0.025 for most values of $\theta_0\leq\pi/4$.  Fig. 6, which contains a
plot of $g_{\rm lim}$ as a function of $M$ for $\theta_0 = 0.1$ radians, shows
that this limit can improve for lower values of $M$, reaching about 0.005 for
$M\approx 300$ GeV.

\section{Discussion}

We now present a discussion of these results. 
First of all, we need to review
the present limits on the couplings and mass of the leptoquarks. Since we
consider specifically \lqs coupling only to third-generation quarks, the direct
limits are rather weak \cite{cdf}. In general, they seem to allow a \lq mass of
about 200 GeV for coupling strengths of the order of electroweak coupling. 
Strong indirect limits may, however, be
obtained especially in our case where the \lq has both left- and right-handed
couplings. Detailed discussion on indirect limits may be found in \cite{limits}.
The most stringent limits come from dipole moments of the electron.
Requiring the contribution to the electron $g-2$ coming from one-loop diagrams
with top and \lq internal lines 
\beq
g_e-2 \approx \frac{\alpha}{2\pi}\,\frac{m_em_t}{M^2}\,\ln \frac{m_t^2}{M^2}
{\rm Re}(g_Rg_L^*)
\eeq
to be less than the experimental uncertainty of $8\times 10^{-12}$ gives
\beq
\frac{ {\rm Re}(g_Rg_L^*) }{ \left(M/{\rm TeV}\right)^2 } < 0.1. 
\eeq 
The limits obtainable from our asymmetry $A_2$ are clearly better than this.

The contribution to the electric dipole moment $d_e$ of the electron from the
same one-loop diagrams is 
\beq
d_e \approx \frac{\alpha}{2\pi}\,\frac{m_t}{M^2}\,\ln \frac{m_t^2}{M^2} 
{\rm Im}(g_Rg_L^*).
\eeq
It is clear that the direct limit obtainable from $A_1$, viz., ${\rm Im} 
(g_Rg_L^*) < 0.025$ for $M=1$ TeV, is nowhere near the much more stringent
limit obtained from the experimental limit of about $10^{-27}$ e cm on $d_e$,
which leads to 
\beq
\frac{ {\rm Im}(g_Rg_L^*)}{ \left(M/{\rm TeV}\right)^2 } < 10^{-6}.
\eeq

In conclusion, we have pointed out azimuthal asymmetries which single out
products of opposite-chirality couplings of scalar leptoquarks and which can provide a
direct test of these in linear collider experiments. Longitudinal beam
polarization, on the other hand, can only put limits on the absolute
values of the left and right chiral couplings. The limit that can be put
on real part of the product of the 
couplings $g_Rg_L^*$ is about 0.025 for reasonable values
of linear collider parameters, and assuming a leptoquark mass of about 1 TeV.
This limit is better than the indirect limit from the $g-2$ of the electron.
It would be interesting to look for the asymmetry $A_2$ if transverse 
polarization is available at a future linear collider.
The imaginary part of this product can in principle be constrained by a
suitable CP-violating azimuthal asymmetry to the same extent. However, 
the much better experimental limit on 
the electric dipole of the electron already makes such a limit redundant.

The discussion here can be extended to scalar leptoquarks transforming as
$(\underline{1},{1 \over 3}, \underline{3}^*)$ representation of the gauge
group with similar conclusions.

\vskip .2cm
\noindent{\bf Acknowledgements} This work is supported by the 
Department of Science and Technology
(DST), Government of India, under project number SP/S2/K-01/2000-II.
The author thanks B. Ananthanarayan and Rohini
Godbole for discussions, and Ananthanarayan also for a careful reading of the manuscript.

\end{document}